\documentclass[11pt]{article}
\usepackage[driver=dvipdfm,a4paper,hmarginratio=1:1,vmarginratio=2:3,totalwidth=15.5cm,totalheight=22.5cm]{geometry}
\usepackage{bm,bbm,epstopdf,epsfig,amsmath,amssymb,amsfonts,colordvi,latexsym,wrapfig,comment,cancel,verbatim,slashed}
\usepackage{graphicx,graphics}
\usepackage[font=md,captionskip=8pt]{subfig}
\usepackage[usenames,dvipsnames]{color}
\usepackage[noadjust]{cite}
\usepackage{xcolor} 
\usepackage[utf8]{inputenc}
\usepackage{setspace}
\usepackage{mathtools}

\usepackage[utf8]{inputenc}
\DeclareMathAlphabet{\pazocal}{OMS}{zplm}{m}{n}

\allowdisplaybreaks

\newcommand{\ra}{\rightarrow}
\newcommand{\be}{\begin{equation}}
\newcommand{\ee}{\end{equation}}
\newcommand{\bea}{\begin{eqnarray}}
\newcommand{\eea}{\end{eqnarray}}

\long\def\symbolfootnote[#1]#2{\begingroup
\def\thefootnote{\fnsymbol{footnote}}\footnote[#1]{#2}\endgroup}
\begin{document} 
\begin{flushright}
\end{flushright}
\bigskip\medskip
\thispagestyle{empty}
\vspace{2cm}

\begin{center}
\vspace{0.5cm}

{\Large \bf Quantum scale invariance in gauge theories

  \bigskip
  and applications to muon production}

 \vspace{1.5cm}
 {\bf M. Wei{\ss}wange} $^{a,\,}$\symbolfootnote[2]{matthias.weisswange@tu-dresden.de},
 {\bf D. M. Ghilencea} $^{b,\,}$\symbolfootnote[1]{dumitru.ghilencea@cern.ch} and 
 {\bf D. St\"ockinger} $^{a,\,}$\symbolfootnote[3]{dominik.stoeckinger@tu-dresden.de}

\bigskip
$^a$ {\small Institut f\"ur Kern- und Teilchenphysik,
  Technische Universit\"at Dresden, 01069  Germany}

\medskip

$^b$ {\small Department of Theoretical Physics, National Institute of Physics 
 \medskip 

 and  Nuclear Engineering (IFIN) Bucharest, 077125 Romania}
\end{center}

\medskip
\begin{abstract}
  We discuss quantum scale invariance in (scale invariant) gauge theories with both
  ultraviolet (UV) and infrared (IR) divergences. Firstly, their BRST invariance
  is checked in two apparently unrelated approaches using a   scale invariant regularisation (SIR).
  These approaches are then shown to be equivalent.
  Secondly, for the Abelian case we discuss both UV and IR quantum corrections
  present in such theories.  We  present the Feynman rules in a form suitable
  for offshell Green functions calculations, together with their one-loop renormalisation.
  This information is  then used for the muon production cross section at one-loop
  in a quantum scale invariant theory. Such a theory contains not only new UV poles but
  also IR poles. While the UV poles  bring new  quantum corrections (in the form of counterterms),  finite or divergent,
  that we compute, it is shown that  the IR poles  do not bring new physics.
  The IR  quantum corrections, both finite and divergent, cancel out
  similarly to the way the IR poles themselves cancel in the traditional approach to IR divergences
  (in the cross section, after summing over virtual and real corrections).
  Hence, the evanescent interactions induced  by the scale-invariant  analytical continuation
  of the SIR scheme do not affect IR physics, as illustrated at one-loop for the muon
  production  ($e^+ e^- \ra \mu^+\mu^-$) cross section. 
  \end{abstract}

\newpage

\section{Introduction}

In this work we study quantum scale invariance and its implications for 
scale invariant gauge theories. An example is the Standard Model with a vanishing Higgs mass
\cite{Bardeen}. Consider such a theory in four dimensions; its quantum corrections are
usually divergent in the ultraviolet  (UV) and possibly in the infrared (IR) as well.
To make the theory well defined,  a  regularisation is needed  and this introduces
a dimensionful parameter i.e. a mass scale, regardless of the regularisation 
(dimensional regularisation (DR), cutoff scheme, etc).  Then the original  theory
in $d=4$ that was classically scale invariant has this symmetry broken explicitly by
the regularisation, e.g. by the  analytical continuation to $d$ dimensions in the DR scheme,
because the DR subtraction scale $\mu$ breaks the classical scale symmetry.

This explicit (anomalous) breaking of scale invariance can be avoided by a manifestly scale
invariant regularisation (SIR), see  \cite{Englert_1976,Armillis_2013, Ghilencea_2016,Shaposhnikov_2009,
  Ghilencea_2018, Ghilencea_2017,Ghilencea_2016_2Loop, gretsch2013dilaton, kugo2020necessity,
  Lalak_2018, Mooij_2019,Matthias_2021}, leading to a quantum scale invariant theory.
One motivation to study such a symmetry  is that it  can naturally
preserve  a classical   hierarchy of  scales generated by field vev's \cite{Shaposhnikov_2009} 
(for an example see \cite{Ghilencea_2017}).
More generally, in gauge theories of scale invariance that include Einstein gravity
\cite{Ghilencea:2018dqd} quantum scale symmetry becomes necessary.

In the SIR scheme an  additional scalar field  $\sigma$ is introduced by the  analytical
  continuation\footnote{unless $\sigma$ is already present in the classical
      theory in $d=4$ as a flat direction in the scalar sector.}  to enforce the
  classical scale symmetry in $d\!=\! 4-2\epsilon$ dimensions. Here  $\sigma$
  is the Goldstone of (global) scale symmetry and will generate the  subtraction scale:
  the ``usual'' DR subtraction scale is replaced by a dilaton dependent
 function $\mu\!\sim\!  z\, \sigma$ ($z$ is a dimensionless parameter)
 that enforces  manifest scale invariance in $d$ dimensions at classical and quantum level.
When $\sigma$ acquires a  non-vanishing vacuum expectation value (vev) $\langle \sigma \rangle$,
the usual subtraction scale $\mu_0\!\sim\!\langle\sigma\rangle$ is generated via
spontaneous breaking  of scale symmetry.
Due to this  there is no anomalous scale symmetry breaking anymore
\cite{Armillis_2013, Englert_1976, Ghilencea_2017, Lalak_2018}.\footnote{Nevertheless,
  the couplings still   run with momentum \cite{Tamarit}, information captured by dimensionless $z$.}
While the analytical continuation to $d$ dimensions preserves scale invariance, note that the
spectrum may then differ from that of the initial  $d=4$ theory by the additional (dynamical)
field $\sigma$.

The above procedure has strong implications at the quantum level. 
After quantum corrections are computed in $d$ dimensions, scale invariant counterterms are
identified and renormalisation is performed. After doing so,  the limit $d \ra 4$ is taken.
Some  quantum corrections depend on the dilaton and these are solely
due to quantum scale invariance. Eventually, the theory can now be expanded about the (large) vev of
the dilaton $\langle\sigma\rangle$ which represents the scale of ``new physics'' (above
which scale invariance is restored). For a large dilaton vev, these corrections
are actually expected to vanish. In this way one has a quantum scale invariant theory that also
recovers the ``usual'' quantum corrected theory (regularised by ``standard''  DR) in the
decoupling limit of the dilaton i.e. at large $\langle\sigma\rangle$. The two theories have,
however,  different UV spectrum, UV symmetry and UV behaviour.

Since the Lagrangian in $d=4-2\epsilon$ dimensions
now depends on $\mu^{\epsilon}(\sigma)$, where
$\mu(\sigma)$ is a mass dimension-one function of $\sigma$, by expanding
in powers of $\epsilon$ one obtains new ``evanescent'' polynomial interactions proportional
to powers of $\epsilon$. Then there are {\it new}, scale invariant quantum
corrections that emerge from evanescent
interactions $\propto\epsilon^n$ hitting a pole $1/\epsilon^m$ arising from loop integrals,
in the cases when $n\!\leq\! m$. If $n=m$, new finite quantum corrections
(which hence have the form of finite counterterms)
are generated, whereas if $n<m$, new poles  are generated leading to new scale
invariant counterterms and quantum corrections, suppressed by powers of $\sigma$.
Such calculations were already performed at one-loop 
\cite{Armillis_2013, Ghilencea_2016, Ghilencea_2017,Shaposhnikov_2009,Mooij_2019,Lalak_2018},
two-loop \cite{Ghilencea_2016_2Loop,Matthias_2021},
and even three-loop level \cite{Ghilencea_2018,gretsch2013dilaton} (for a review see
\cite{kugo2020necessity}).

Quantum scale invariance was studied in the past especially in the scalar sector,
whereas gauge theories were less studied  except  in \cite{Armillis_2013,Mooij_2019,Ghilencea_2017}
and this motivated this work. In particular, one aspect that was overlooked is
the BRST invariance of these theories following their analytical continuation to $d\!=\! 4-2\epsilon$
in a manifestly scale invariant way.  Secondly, the scale invariant analytical
continuation for gauge theories can be performed in two different ways, apparently unrelated,
which are not known to be equivalent. We  show  that both
approaches lead to the same $d$ dimensional, BRST invariant Lagrangian and are thus equivalent.

The third and strongest motivation of this study is related to the fact
that (scale invariant) gauge theories have not only UV but also IR divergences which were
not yet discussed in the context of quantum scale invariance.
The SIR scheme  based on the DR scheme can actually handle the IR
divergences as well. A question is whether quantum scale invariance
leads to new physical effects related to IR divergences.
We show  that, while the  UV poles  bring new  quantum corrections due to  quantum scale invariance,
the IR poles  do not bring any new physics. The new infrared quantum corrections, be they
{\it finite or divergent}, cancel out,  similar to the way the IR poles themselves cancel in the traditional
approach to IR divergences (at the level of cross sections after summing over virtual and real corrections).
Thus, ``evanescent'' interactions from the SIR do not impact IR physics.
This cancellation is a strong consistency check for quantum scale invariant gauge theories.
We illustrate this in detail
with a one-loop calculation of the cross section of muon production,
$e^{-} \, e^{+} \rightarrow \mu^{-} \, \mu^{+}$; this is evaluated and analysed 
in the framework of quantum electrodynamics extended by the dilaton which
enforces quantum scale invariance.

The plan of the paper is as follows: Section~\ref{2} discusses the BRST invariance
in  SIR schemes for general gauge theories.
Since gauge theories also have  IR poles, we study these in an
Abelian  gauge theory, derive Feynman rules (Appendix~\ref{Feynman-rules})
and its renormalisation  (Appendix~\ref{Renormalisation}).
This information is then used in Section~\ref{3} that presents the muon production cross section and the
effect of UV and IR corrections, with technical details in
Appendix~\ref{Feynman} and \ref{D}. Our conclusions are presented in the last section.

\section{Quantum scale invariance in gauge theories}\label{2}

We check the BRST invariance in general gauge theories in two apparently unrelated
approaches in SIR schemes and show the equivalence of their results. Further,
gauge theories  have  not only UV poles, but also IR ones, the study of which is
restricted to the  Abelian case, in the SIR scheme.   We thus consider an Abelian case
(quantum electrodynamics extended by the dilaton), for which we give
the Feynman diagrams and one loop renormalisation in a manifestly scale invariant form
suitable for computer implementation.

\subsection{Non-Abelian theories and BRST}\label{Sec21}

Consider the Lagrangian of a  scale invariant \(SU(N)\) gauge theory in $d=4$ given by
\begin{equation} \label{L01}
    \begin{aligned}
      \pazocal{L} =
      \pazocal{L}_{cl}
      + \pazocal{L}_{\mathrm{GF}}
      + \pazocal{L}_{\mathrm{Ghost}}
    \end{aligned}
\end{equation}
\begin{equation}
    \label{L02}
    \begin{aligned}
      \pazocal{L}_{cl} = &- \frac{1}{4} \, F_{\mu \nu}^{a} \, F^{a, \mu \nu}
      + i \, \overline{\psi}_{i} \left( \delta_{i j} \, \slashed{\partial}
        - i \, g \, \slashed{G}^{a} \, T_{i j}^{a}\right) \psi_{j}
    \end{aligned}
\end{equation}
\begin{equation} \label{L03}
  \begin{aligned}
        \pazocal{L}_{\mathrm{GF}} = - B^{a} \, \partial^{\mu} G_{\mu}^{a} + \frac{\xi}{2} \, B^{a} \, B^{a}
    \end{aligned}
  \end{equation}
  \begin{equation} \label{L04}
    \begin{aligned}
        \pazocal{L}_{\mathrm{Ghost}} = \partial^{\mu} \overline{c}^{a} \, D_{\mu}^{a c} c^{c}
    \end{aligned}
\end{equation}
{where \(B^{a}\), \(c^{a}\) and \(\overline{c}^{a}\) are Nakanishi-Lautrup
  fields, ghosts and anti-ghosts, respectively.}

Apparently, there are  two different ways to  implement a scale invariant regularisation:

\vspace{0.6cm}\noindent
{\bf (a) Analytical continuation to $d=4-2\epsilon$ by rescaling the gauge fields}

\vspace{0.5cm}
\noindent
First, rescale the gauge fields in $d=4$ by a factor of the gauge coupling,
\begin{equation} \label{Rescaling1}
    \begin{aligned}
        G_{\mu}^{a} &\longrightarrow \hat{G}_{\mu}^{a} = g \, G_{\mu}^{a}.
    \end{aligned}
\end{equation}

\noindent
Since the gauge parameter \(\beta^{a}\) also needs to be rescaled, the ghost \(c^{a}\) is
analogously rescaled, as \(\beta^{a}(x) = \theta \, c^{a}(x)\), for some Grassmann number \(\theta\). 
Further, it is convenient to rescale the anti-ghost \(\overline{c}^{a}\) as well in order
to obtain a ghost term similar to (\ref{L04}). Thus,
\begin{equation} \label{Rescaling2}
    \begin{aligned}
      c^{a} \longrightarrow \hat{c}^{a} = g \, c^{a}, \hspace{0.75cm} \overline{c}^{a}
      \longrightarrow \hat{\overline{c}}^{a} = \frac{1}{g} \, \overline{c}^{a}.
    \end{aligned}
\end{equation}
\noindent
This leads to
\begin{equation}
  \label{L1}
    \begin{aligned}
      \pazocal{L} = &- \frac{1}{4 \, g^{2}} \, \hat{F}_{\mu \nu}^{a} \, \hat{F}^{a, \mu \nu}
      + i \, \overline{\psi} \, \hat{\slashed{D}} \, \psi
      - \frac{1}{g} \, B^{a} \, \partial^{\mu} \hat{G}_{\mu}^{a}
        + \frac{\xi}{2} \, B^{a} \, B^{a} + \partial^{\mu} \hat{\overline{c}}^{a} \, \hat{D}_{\mu}^{a c} \hat{c}^{c},
    \end{aligned}
\end{equation}
\noindent{where the rescaled field strength tensor and covariant
derivative do not explicitly depend on the gauge coupling,}
\begin{align} \label{L11}
  \hat{F}_{\mu \nu}^{a} &= \partial_{\mu} \hat{G}_{\nu}^{a} -
                          \partial_{\nu} \hat{G}_{\mu}^{a} + f^{abc} \,
                          \hat{G}_{\mu}^{b} \, \hat{G}_{\nu}^{c}\\[5pt]
  \hat{D}_{\mu} \, \psi_i &= \left(\delta_{ij} \partial_{\mu} - i \, \hat{G}_{\mu}^{a} \,
                          T^{a}_{ij} \right) \psi_j\\[5pt]
  \hat{D}_{\mu}^{a c} \hat{c}^{c} &= \left( \delta^{a c} \, \partial_{\mu}
                                    + f^{a b c} \, \hat{G}_{\mu}^{b} \right) \hat{c}^{c}.
\end{align}

\medskip
Lagrangian (\ref{L1})
can now be analytically continued to $d = 4 - 2 \epsilon$ in a scale invariant way.
This is achieved by the addition of a {\it dynamical} scalar field $\sigma$
to the spectrum of the initial classical theory in $d=4$,  playing the role of a dilaton. 
When $\sigma$ acquires a vev, the subtraction scale is generated dynamically.
Hence, the analytical continuation to $d$ dimensions modifies the spectrum
of the initial classical theory in $d=4$ by an extra degree of freedom,
but scale invariance is maintained in $d$ dimensions (and at the quantum level).
This is the ``cost'' of implementing quantum scale invariance. 
The Lagrangian becomes
\medskip
\begin{equation} \label{L2}
\begin{aligned}
  \pazocal{L}^{(d)} = &- \frac{1}{4 \, g^{2}} \,
                        \mu^{- 2 \epsilon}(\sigma) \, \hat{F}_{\mu \nu}^{a}
                        \, \hat{F}^{a, \mu \nu} + i \, \overline{\psi} \,
                        \hat{\slashed{D}} \, \psi + \frac{1}{2} \left( \partial_{\mu}
                        \sigma \right) \left( \partial^{\mu} \sigma \right)\\
                      &- \frac{1}{g} \, \mu^{- \epsilon}(\sigma) \, B^{a} \, \partial^{\mu} \hat{G}_{\mu}^{a}
                        + \frac{\xi}{2} \, B^{a} \, B^{a} + \partial^{\mu} \hat{\overline{c}}^{a} \, \hat{D}_{\mu}^{a c} \hat{c}^{c}.
\end{aligned}
\end{equation}
%
The mass dimensions of the fields and couplings are
\begin{equation} \label{massdims1}
    \begin{aligned}
        &[\hat{G}_{\mu}^{a}] = 1,
        &\hspace{0.5cm}
        &[\psi] = \frac{3}{2} - \epsilon,
        &\hspace{0.5cm}
        &[\sigma] = 1 - \epsilon,
        &\hspace{0.5cm}
        &[g] = 0,\\
        &[\hat{c}^{a}] = 0,
        &\hspace{0.5cm}
        &[\hat{\overline{c}}^{a}] = 2 - 2 \epsilon,
        &\hspace{0.5cm}
        &[B^{a}] = 2 - \epsilon,
        &\hspace{0.5cm}
        &[\xi] = 0
    \end{aligned}
  \end{equation}
\noindent
and  \([\hat{\beta}^{a}] = 0\), \([\theta] = 0\).
Further, integrating out $B^a$ in (\ref{L2}), then
\begin{equation}
    \begin{aligned}
      \pazocal{L}^{(d)} = &- \frac{1}{4 \, g^{2}} \, \mu^{- 2 \epsilon}(\sigma)
      \, \hat{F}_{\mu \nu}^{a} \, \hat{F}^{a, \mu \nu} + i \, \overline{\psi}
      \, \hat{\slashed{D}} \, \psi + \frac{1}{2} \left( \partial_{\mu}
        \sigma \right) \left( \partial^{\mu} \sigma \right)\\
      &- \frac{1}{2 \, \xi \, g^{2}} \, \mu^{- 2 \epsilon}(\sigma)
      \left( \partial^{\mu} \hat{G}_{\mu}^{a} \right)^{2} + \partial^{\mu} \hat{\overline{c}}^{a} \,
      \hat{D}_{\mu}^{a c} \hat{c}^{c}.
    \end{aligned}
\end{equation}

\medskip
\noindent
In general, the BRST transformations are given by
\begin{equation} \label{brst1}
    \begin{aligned}
        &\psi_{i} \longmapsto \psi_{i} + \delta \psi_{i},
        &\hspace{0.5cm}
        &G_{\mu}^{a} \longmapsto G_{\mu}^{a} + \delta G_{\mu}^{a},
        &\hspace{0.5cm}
        &c^{a} \longmapsto c^{a} + \delta c^{a},\\[4pt]
        &\overline{\psi}_{i} \longmapsto \overline{\psi}_{i} + \delta \overline{\psi}_{i},
        &\hspace{0.5cm}
        &B^{a} \longmapsto B^{a} + \delta B^{a},
        &\hspace{0.5cm}
        &\overline{c}^{a} \longmapsto \overline{c}^{a} + \delta \overline{c}^{a},\qquad
        \sigma \longmapsto \sigma,
    \end{aligned}
\end{equation}

\medskip\noindent
where \(\sigma\) transforms trivially under BRST symmetry.
One finds the following $d$ dimensional BRST transformations:
\bea \label{brst2}
    \begin{aligned}
      \delta \psi_{i} &= \theta \pazocal{Q} \psi_{i} = i \, \theta \, \hat{c}^{a} \, T_{i j}^{a} \, \psi_{j},\\[5pt]
              \delta \overline{\psi}_{i} &= \theta \pazocal{Q} \overline{\psi}_{i} =
        - i \, \theta \, \hat{c}^{a} \, \overline{\psi}_{j} \,  T_{j i}^{a}\\[5pt]
        \delta \hat{G}_{\mu}^{a} &= \theta \pazocal{Q} \hat{G}_{\mu}^{a} =
        \theta \hat{D}_{\mu}^{a c} \hat{c}^{c} = \theta \, \partial_{\mu} \hat{c}^{a}
        + \theta \, f^{a b c} \, \hat{G}_{\mu}^{b} \, \hat{c}^{c}\\[5pt]
        \delta \hat{c}^{a} &= \theta \pazocal{Q} \hat{c}^{a} = - \frac{1}{2} \,
        \theta \, f^{a b c} \, \hat{c}^{b} \, \hat{c}^{c}\\
        \delta \hat{\overline{c}}^{a} &= \theta \pazocal{Q} \hat{\overline{c}}^{a}
        = - \frac{\theta}{g} \, \mu^{-\epsilon}(\sigma) \, B^{a} =
        - \frac{\theta}{\xi \, g^{2}} \, \mu^{- 2 \epsilon}(\sigma) \, \partial^{\mu} \hat{G}_{\mu}^{a}\\[5pt]
        \delta B^{a} &= \theta \pazocal{Q} B^{a} = 0,
    \end{aligned}
\eea
\noindent
where the first three transformations are given by the infinitesimal gauge transformations,
as usual, using \(\hat{\beta}^{a}(x) = \theta \, \hat{c}^{a}(x)\).

It is straightforward to show that the BRST operator \(\pazocal{Q}\) that generates the BRST
transformations in (\ref{brst2})
is nilpotent as it should be, i.e. \(\pazocal{Q}^{2} = 0\). Furthermore, the $d$ dimensional Lagrangian
(\ref{L2}) is indeed  BRST invariant, i.e. invariant under transformations (\ref{brst2}), which can be shown
by direct calculation.

Although a detailed all-orders investigation is beyond the scope of the
present work, let us briefly comment on the role of the BRST invariance
in the case of renormalization of the considered gauge theories, 
which are non-renormalizable due to quantum scale invariance.
We expect BRST invariance to be crucial for the consistency of the
quantized theory, just as in ordinary gauge theories, which are
strictly renormalizable and thus involve only a finite number of
operators in the Lagrangian.

At higher orders, we expect that BRST invariance will control
the structure of possible UV divergences, and required counterterms
are expected to be BRST invariant as well, even if there will be
infinitely-many of these. To understand this, note that the
situation  is similar to the  general analysis in the review
\cite{Barnich_2000}, although our particular case with a quantum scale symmetry
is not covered in this reference.
However, note that in our theory the dilaton transforms trivially under BRST transformations,
i.e. behaves like a BRST singlet; quantum scale invariance imposes additional 
constrains on the theory and on the structure of its UV-counterterms.
In the broken phase of quantum scale symmetry and  the decoupling limit
of the dilaton i.e. $\langle\sigma\rangle \rightarrow \infty$, the higher dimensional operators
vanish (since they are suppressed by the vev of the dilaton) and one
recovers a renormalizable theory where traditional results and BRST constraints 
of gauge theories apply.  These constraints extend however to the symmetric
phase of the theory since the quantum scale symmetry is broken only spontaneously.
Ultimately, we expect an all-orders Slavnov-Taylor identity to hold,
establishing physical properties such as unitarity and gauge-parameter
independence of the physical S-matrix, cf.\ \cite{Kugo_1977,Kugo_1979}.
The unitarity of the physical S-matrix was also verified \cite{Oda}
in more general theories  where scale symmetry is {\it gauged}
(e.g. \cite{Ghilencea:2018dqd}) and quantum scale invariance
is thus automatic.

\vspace{1cm}

\noindent
{\bf (b) A different approach to $d=4-2\epsilon$}

\bigskip
\noindent
There is a second approach to a scale invariant regularisation of gauge theories
that  is somewhat  ``geometrical''.
First, by analytical continuation to \(d\) dimensions, the covariant derivative changes
\begin{equation} \label{CovD2}
    \begin{aligned}
      D_{\mu} \longrightarrow \widetilde{D}_{\mu} = \partial_{\mu} -
      i \, g \, \mu^{\epsilon}(\sigma) \, G_{\mu}^{a} \, T^{a}.
    \end{aligned}
\end{equation}

\medskip\noindent
The field strength, regarded as the curvature tensor of the internal coordinates, is then
\medskip
\begin{equation} \label{F2}
    \begin{aligned}
      \widetilde{F}_{\mu\nu} = \widetilde{F}_{\mu\nu}^{a} \, T^{a} =
      \frac{i}{g \, \mu^{\epsilon}(\sigma)} \, \Big[ \widetilde{D}_{\mu},\widetilde{D}_{\nu} \Big].
    \end{aligned}
\end{equation}
Evaluating the commutator leads to
\begin{equation} \label{F}
    \begin{aligned}
      \widetilde{F}_{\mu\nu}^{a} = \partial_{\mu} G_{\nu}^{a} - \partial_{\nu} G_{\mu}^{a}
      + g \, \mu^{\epsilon}(\sigma) \, f^{abc} \, G_{\mu}^{b} \, G_{\nu}^{c} + \epsilon \, \mu^{-1}(\sigma) \,
      \frac{\partial \mu}{\partial \sigma} \, \Big(\partial_{\mu} \sigma \, G_{\nu}^{a}
      - \partial_{\nu} \sigma \, G_{\mu}^{a} \Big),
    \end{aligned}
\end{equation}
so $\tilde F$ has received an evanescent correction ($\propto\epsilon$).
The Lagrangian $\pazocal{L}_{cl}^{(d)}$ in $d=4-2\epsilon$ is then 
\medskip
\begin{equation} \label{LL}
    \begin{aligned}
      \pazocal{L}_{cl}^{(d)} = &- \frac{1}{4} \, \widetilde{F}_{\mu \nu}^{a} \,
      \widetilde{F}^{a, \mu \nu} + i \, \overline{\psi}_{i} \left( \delta_{i j}
        \, \slashed{\partial} - i \, g \, \mu^{\epsilon}(\sigma) \, \slashed{G}^{a}
        \, T_{i j}^{a}\right) \psi_{j} + \frac{1}{2} \left( \partial_{\mu} \sigma \right)
      \left( \partial^{\mu} \sigma \right).
    \end{aligned}
\end{equation}

\medskip
\noindent
Similarly, the combination $\mu^{\epsilon}(\sigma) \, G_{\mu}^{a}$ is
the one which determines the gauge fixing and ghost Lagrangian as well
as the BRST transformations.\footnote{The BRST transformations also
involve the combination $\mu^{\epsilon}(\sigma) \, c^{a}$ in an
essential way, see later, eq. (\ref{brst3}).}
The gauge fixing and ghost Lagrangian is
\medskip
\begin{equation} \label{L3}
    \begin{aligned}
      \pazocal{L}_{\mathrm{GF}}^{(d)} + \pazocal{L}_{\mathrm{Ghost}}^{(d)}
      &= \pazocal{Q} \left[ - \overline{c}^{a} \left( \frac{\xi}{2} \, B^{a}
          - \partial^{\mu} G_{\mu}^{a}
          - \epsilon \, \mu^{-1}(\sigma) \,
          \frac{\partial \mu}{\partial \sigma} \, \partial^{\mu} \sigma \, G_{\mu}^{a} \right) \right]\\
      &= \frac{\xi}{2} \, B^{a} \, B^{a} - B^{a} \, \partial^{\mu} G_{\mu}^{a}
      + \partial^{\mu} \overline{c}^{a} \, \widetilde{D}_{\mu}^{a c} c^{c}\\
      &\hspace{0.44cm} - \epsilon \, \mu^{-1}(\sigma) \, \frac{\partial \mu}{\partial \sigma} \, \partial^{\mu} \sigma \Big[ B^{a} \, G_{\mu}^{a}
      + \overline{c}^{a} \, \widetilde{D}_{\mu}^{a c} c^{c}
      - \partial_{\mu} \overline{c}^{a} \, c^{a} \Big]\\
      &\hspace{0.44cm} - \epsilon^{2} \, \mu^{-2}(\sigma) \left(\frac{\partial \mu}{\partial \sigma}\right)^{2}
      \partial^{\mu} \sigma \, \partial_{\mu} \sigma \, \overline{c}^{a} c^{a},
    \end{aligned}
\end{equation}
\noindent{where}
\begin{equation} \label{CovD}
    \begin{aligned}
      \widetilde{D}_{\mu}^{a c} = \delta^{a c} \, \partial_{\mu}
      + g \, \mu^{\epsilon}(\sigma) \, f^{a b c} \, G_{\mu}^{b}.
    \end{aligned}
\end{equation}

\medskip
\noindent
It can be seen that the gauge fixing and the ghost Lagrangian also acquire
purely evanescent corrections ($\propto \epsilon$), analogously to the kinetic term of the gauge field,
eq.(\ref{F}). Consequently,
the \(d\) dimensional Lagrangian of the considered \(SU(N)\) gauge theory is given by
\medskip
\begin{equation} \label{whatever}
    \begin{aligned}
        \pazocal{L}^{(d)} &= \pazocal{L}_{cl}^{(d)} + \pazocal{L}_{\mathrm{GF}}^{(d)} + \pazocal{L}_{\mathrm{Ghost}}^{(d)}.
    \end{aligned}
\end{equation}

\noindent
The \(d\) dimensional BRST transformations in this approach are found to be:
\begin{equation}
\label{brst3}
   \begin{aligned}
     \delta \psi_{i} &= \theta \pazocal{Q} \psi_{i} = i \, \theta \, g \,
     \mu^{\epsilon}(\sigma) \, c^{a} \, T_{i j}^{a} \, \psi_{j}\\[5pt]
\delta \overline{\psi}_{i} &= \theta \pazocal{Q} \overline{\psi}_{i} =
- i \, \theta \, g \, \mu^{\epsilon}(\sigma) \, c^{a} \, \overline{\psi}_{j} \,  T_{j i}^{a}\\[5pt]
        \delta G_{\mu}^{a} &= \theta \pazocal{Q} G_{\mu}^{a} =
        \theta \, \widetilde{D}_{\mu}^{a c} c^{c} + \epsilon \, \theta \, \mu^{-1}(\sigma) \, \frac{\partial \mu}{\partial \sigma} \, \partial_{\mu} \sigma \, c^{a}\\
        &\hspace{1.64cm} = \theta \, \partial_{\mu} c^{a}
        + \theta \, g \, \mu^{\epsilon}(\sigma) \, f^{a b c} \, G_{\mu}^{b} \, c^{c}
        + \epsilon \, \theta \, \mu^{-1}(\sigma) \, \frac{\partial \mu}{\partial \sigma} \,
        \partial_{\mu} \sigma \, c^{a}\\
        \delta c^{a} &= \theta \pazocal{Q} c^{a} =
        - \frac{1}{2} \, \theta \, g \, \mu^{\epsilon}(\sigma) \, f^{a b c} \, c^{b} \, c^{c}\\
        \delta \overline{c}^{a} &= \theta \pazocal{Q} \overline{c}^{a} = - \theta \, B^{a}\\[5pt]
        \delta B^{a} &= \theta \pazocal{Q} B^{a} = 0,
 \end{aligned}
\end{equation}
\noindent
where the first three BRST transformations in (\ref{brst3})
are again given by the gauge transformation, as usual, using $\beta^a(x)=\theta\,c^a(x)$.
  The new, evanescent correction to the gauge field BRST transformation \(\delta G_{\mu}^{a}\)
  has its origin in the dilaton dependent function \(\mu(\sigma)\). In particular, this correction
  originates from \(\partial_{\mu} U\), with \(U = \exp(i g \mu^{\epsilon}(\sigma) \beta^{a} T^{a})\),
  in the derivation
  of \(\delta G_{\mu}^{a}\). Again, the BRST operator \(\pazocal{Q}\),
  generating the BRST transformations in (\ref{brst3}),
  is nilpotent, \(\pazocal{Q}^{2} = 0\). Moreover, while the gauge
  fixing and ghost Lagrangian may be written as a \(\pazocal{Q}\)-exact
  term, as in (\ref{L3}), the BRST invariance of \(\pazocal{L}_{cl}^{(d)}\) in
  (\ref{LL})
  can again be shown by explicit calculation. Thus, $\pazocal{L}^{(d)}$ in
  (\ref{whatever}) is BRST invariant under (\ref{brst3}).

The mass dimensions of the fields and parameters in (\ref{whatever}) are
\begin{equation} \label{massdim2}
    \begin{aligned}
        &[G_{\mu}^{a}] = 1 - \epsilon,
        &\hspace{0.5cm}
        &[\psi] = \frac{3}{2} - \epsilon,
        &\hspace{0.5cm}
        &[\sigma] = 1 - \epsilon,
        &\hspace{0.5cm}
        &[g] = 0,\\
        &[c^{a}] = - \epsilon,
        &\hspace{0.5cm}
        &[\overline{c}^{a}] = 2 - \epsilon,
        &\hspace{0.5cm}
        &[B^{a}] = 2 - \epsilon,
        &\hspace{0.5cm}
        &[\xi] = 0
    \end{aligned}
\end{equation}
\noindent{and \([\beta^{a}] = - \epsilon\), \([\theta] = 0\).}
Compared to the previous approach, eq.(\ref{massdims1}),
notice the different dimensions of ghost/anti-ghost, gauge field and $\hat\beta^a$.


\subsection{Equivalence of the two approaches}

The two approaches of (a) and (b) and their corresponding Lagrangians, eqs.(\ref{L2}) and
(\ref{whatever}),  must be equivalent.
To see this,
start with (\ref{L2}), then the gauge coupling \(g\) and
the subtraction function \(\mu^{\epsilon}(\sigma)\) can be factored
out from the fields \(\hat{G}_{\mu}^{a}\), \(\hat{c}^{a}\) and \(\hat{\overline{c}}^{a}\)
leading to  Lagrangian (\ref{whatever}).
Conversely, starting from eq. (\ref{whatever}), \(g\) and \(\mu^{\epsilon}(\sigma)\)
can be absorbed into \(G_{\mu}^{a}\), \(c^{a}\) and \(\overline{c}^{a}\). Then,
additional terms coming from commuting derivatives and the subtraction function
must be taken into account by subtracting them, leading to eq. (\ref{L2}).
Moreover, the same equivalence holds true for the BRST transformations in (\ref{brst2})
and (\ref{brst3}).

For practical calculations, Lagrangian (\ref{L2}) takes a more convenient form than (\ref{whatever})
as it avoids the evanescent corrections to the gauge kinetic term and to the gauge
fixing and ghost Lagrangians, eqs. (\ref{F}) and (\ref{L3}), respectively.

Further, in approach (a), \emph{after} the theory was analytically continued to
$d = 4 - 2 \epsilon$ a second field redefinition can be applied where the
dimensionless gauge coupling \(g\) is ``extracted''
from  the gauge field
by replacing\footnote{In  the next section we use approach (a)  in terms of
  $\overline{G}_{\mu}^{a}$ but the ``overbar'' is not displayed, for simplicity.}
  \begin{equation} \label{def1}
            \begin{aligned}
 \hat{G}_{\mu}^{a} &= g \, \mu^{\epsilon}(\sigma) \, G_{\mu}^{a} = g \, \overline{G}_{\mu}^{a}.
            \end{aligned}
          \end{equation}
The mass dimension of the gauge field remains $[\overline{G}_{\mu}^{a}] = 1$.
However, the \(d\) dimensional Lagrangian in case (a) then looks more similar
to (\ref{L01}) w.r.t. the gauge couplings,
which is  useful in the presence of mixing between the gauge fields as in the SM.


\subsection{Abelian theories: QED + dilaton}

In (scale invariant) gauge theories, one encounters
not only UV but also IR poles which remains true in our SIR scheme.
We  consider here the simpler case of Abelian gauge theories\footnote{The case of IR divergences
  in non-Abelian case is significantly more difficult.}
to illustrate how both UV and IR poles are handled in this scheme in applications (Section~\ref{3}).
 Consider then quantum electrodynamics in $d=4-2\epsilon$ dimensions, analytically
continued in a scale invariant way. This means the action is  ``upgraded'' to include the
additional dilaton field, as already discussed. In this  Abelian case,
the Faddeev-Popov ghosts completely decouple and are not shown.
Using approach (a) discussed earlier, the Lagrangian becomes\footnote{
 Note that we include an additional tree-level dilaton Yukawa interaction term. This term is gauge and scale invariant, but it was
 not considered in the discussion above in Section \ref{Sec21}. Furthermore, one can in principle
 also have a $\lambda\sigma^{4}$-coupling, which is, however, not
 considered here.}
\medskip
\begin{equation}\label{QSIQED}
    \begin{aligned}
        \pazocal{L}^{(d)} = &- \frac{1}{4} \, \mu^{-2\epsilon}(\sigma) \, F_{\mu \nu} \, F^{\mu \nu} + i \, \overline{\psi}_{f} \left( \slashed{\partial} - i \, e \, Q_{f} \, \slashed{A} \right) \psi_{f} + \frac{1}{2} \left( \partial_{\mu} \sigma \right) \left( \partial^{\mu} \sigma \right)\\
        &- y_{f} \, \mu^{\epsilon}(\sigma) \, \sigma \, \overline{\psi}_{f} \, \psi_{f} - \frac{1}{2 \, \xi} \, \mu^{-2\epsilon}(\sigma) \left(\partial^{\mu} A_{\mu}\right)^{2},
    \end{aligned}
\end{equation}
\noindent
The Lagrangian is scale invariant in $d$ dimensions. Since scale invariance is broken
spontaneously, this does not affect the ultraviolet behaviour of the theory.
In principle, one can work either in the symmetric phase
or in the spontaneously broken phase obtained by an expansion about the vev of the dilaton
but keeping all terms in such an expansion. Then the counterterms and thus the
quantum corrections are not affected.

For practical purposes and for Feynman diagrams derivation,
$\pazocal L^{(d)}$ is expanded first in powers of $\epsilon$
with every order in $\epsilon$ (and loop order) being manifestly scale invariant. Subsequently,
one can express the dilaton in terms of its fluctuations about its non-zero vev
$\langle\sigma\rangle\equiv w$, $\sigma=w+\mathfrak{D}$ and expand in powers of
$\eta=\mathfrak{D}/w$ \cite{Ghilencea_2018}; scale invariance is maintained by
including all terms of this second expansion.
Accordingly, for the function $\mu(\sigma) = z \, \sigma^{\frac{1}{1-\epsilon}}$ in $\pazocal L^{(d)}$ one has
\begin{equation}
    \begin{aligned}
      \mu^{k \epsilon}(\sigma) &= \mu_{0}^{k \epsilon} \bigg[ 1 +
      \epsilon k \Big( \eta - \frac{1}{2} \eta^{2} + \frac{1}{3} \eta^{3}
      - \frac{1}{4} \eta^{4} + \mathcal{O}\big(\eta^{5}\big) \Big)\\
      &\hspace{0.97cm}+ \epsilon^{2} k \Big( \eta - \frac{1-k}{2} \eta^{2}
      + \frac{2-3k}{6} \eta^{3} - \frac{6-11k}{24} \eta^{4}
      + \mathcal{O}\big(\eta^{5}\big) \Big) + \mathcal{O}\big(\epsilon^{3}\big) \bigg],
    \end{aligned}
  \end{equation}
for  integer $k$.  With this, the Lagrangian becomes\footnote{Note that the counterterm
  Lagrangian (see Appendix~\ref{Renormalisation}) must be expanded accordingly.}
\begin{equation} \label{QSIQED-taylor}
    \begin{aligned}
      \pazocal{L}^{(d)} = &- \frac{1}{4} \, \mu^{-2\epsilon}_{0} \, F_{\mu \nu} \, F^{\mu \nu}
      + i \, \overline{\psi}_{f} \left( \slashed{\partial} - i \, e \, Q_{f} \, \slashed{A} \right) \psi_{f}
      + \frac{1}{2} \left( \partial_{\mu} \mathfrak{D} \right) \left( \partial^{\mu} \mathfrak{D} \right)\\
        &- \mu^{\epsilon}_{0} \, y_{f} \, w \, \overline{\psi}_{f} \, \psi_{f}
        - \mu^{\epsilon}_{0} \, \Big[1 + \epsilon + \epsilon^{2}
        + \pazocal{O}\left(\epsilon^{3}\right)\Big] \, y_{f} \, \mathfrak{D} \, \overline{\psi}_{f} \, \psi_{f}\\
        &- \frac{1}{2 \, \xi} \, \mu^{-2\epsilon}_{0} \left(\partial^{\mu} A_{\mu}\right)^{2}
        - \mu^{\epsilon}_{0} \, \Big[\epsilon \left(1 + 2 \epsilon\right)
        + \pazocal{O}\left(\epsilon^{3}\right)\Big] \, \frac{y_{f}}{2 w} \, \mathfrak{D}^{2} \, \overline{\psi}_{f} \, \psi_{f}\\
        &+ \mu^{\epsilon}_{0} \, \Big[\epsilon \left(1 + \epsilon\right)
        + \pazocal{O}\left(\epsilon^{3}\right)\Big] \, \frac{ y_{f}}{6 w^{2}} \, \mathfrak{D}^{3} \, \overline{\psi}_{f} \, \psi_{f}\\
        &+ \mu^{-2\epsilon}_{0} \, \Big[\epsilon \left(1 + \epsilon\right)
        + \pazocal{O}\left(\epsilon^{3}\right)\Big] \, \frac{\mathfrak{D}}{w} \, \left[ \frac{1}{2} F_{\mu \nu} \, F^{\mu \nu}
          +  \frac{1}{\xi} \left(\partial^{\mu} A_{\mu}\right)^{2}\right] + \cdots
    \end{aligned}
\end{equation}

\noindent
While every loop order (encoded in powers of $\epsilon$) keeps manifest scale invariance,
any truncation of the subsequent  expansion in powers of $\eta$ breaks this symmetry.
Nevertheless,  the Lagrangian as shown in (\ref{QSIQED-taylor}) is useful
for deriving the Feynman rules that  formally keep all terms in the expansion about $w$.
These are presented in Appendix~\ref{Feynman-rules} and used in Section~\ref{3}.

For completeness, the renormalisation of (\ref{QSIQED}) at one-loop using (\ref{QSIQED-taylor})
is shown in Appendix~\ref{Renormalisation}. Notice that, unlike in previous literature,
$\sigma$ undergoes a wavefunction renormalisation already at one-loop, due to Yukawa interactions.


\section{Application: muon production at one-loop}\label{3}

In this section we use the formalism of the previous section for the Abelian case and
discuss the effect of both UV and IR poles in the SIR scheme. 
We illustrate this for the muon production  at one-loop level
based on Lagrangian (\ref{QSIQED}), for \(3\)
fermion flavours, i.e. \(f \in \{e^{-}, \mu^{-}, \tau^{-}\}\), \(N_{f} = 3\)
and $Q_{f} = - 1$.\footnote{In principle, the
  corrections that we find can be used to set bounds on the scale of ``new physics''
  represented by $\langle\sigma\rangle$,
  provided one considered a full scale invariant SM
  (see \cite{Ghilencea_2017, Matthias_2021} for the Lagrangian and $V_{\mathrm{eff}}$).}

\subsection{General considerations}
\label{Chapter 3 Muon Production}

The scattering process
\(e^{-} \, e^{+} \rightarrow \mu^{-} \, \mu^{+}\) is computed
in the \(\overline{\mathrm{MS}}\)-scheme and Feynman gauge \(\xi = 1\).
We work in the approximation  $m_f^2\ll s$, where $s$ is the center-of-mass energy, i.e.\ the fermions are essentially
massless.\footnote{Practically, this scenario is realised by setting \(m_{f}=0\)
  in the free Lagrangian, i.e. in (A-1), but
  keeping \(y_{f}\) and \(w\) non-zero in the interaction Lagrangian, i.e. in
  (A-2) and (A-3), despite
  $m_{f} = \mu_{0}^{\epsilon} \, y_{f} \, w$.}  This approximation displays a more
interesting IR-divergence structure than the massive case as it contains not only
a simple pole but also a second order  pole in \(\epsilon_{\mathrm{IR}}\). Hence,
this leads to both new finite and also new divergent quantum corrections induced
by evanescent interactions; the latter emerge when a term \(\sim \epsilon\) ``meets''
a second order pole in \(\epsilon_{\mathrm{IR}}\).

For simplicity, only the one-loop muon vertex corrections contributing to
the above scattering process are considered, and thus only final state
real emission needs to be taken into account in order to cancel the IR-divergences
at the level of cross sections. Despite this simplification, the result admits
nonetheless the same general structure as the full one-loop result, meaning that
it contains a simple and a second order pole in \(\epsilon_{\mathrm{IR}}\) as well
as new finite and divergent quantum corrections, see below.
Consequently, for the present conceptual study, it is sufficient to restrict
oneself to one-loop muon vertex corrections.
All  necessary Feynman diagrams,  contributing to virtual and
real muon corrections up to the one-loop level, are  found in Appendix \ref{Feynman}.

\subsection{The cross section at one-loop}

The scattering amplitude and the cross section were
calculated in the above set up by using standard methods and \texttt{Mathematica}
\cite{Wolfram}. In particular, all Feynman diagrams have been generated using
\texttt{FeynArts} \cite{Hahn_2001}, 
with \texttt{FeynArts} model files 
generated by \texttt{FeynRules} \cite{Alloul_2014, Christensen_2009}. The generated
Feynman diagrams and their amplitudes have been computed using \texttt{FeynCalc}
\cite{MERTIG1991345, Shtabovenko_2020, Shtabovenko_2016} and \texttt{Package-X}
\cite{Patel_2015}, connected to \texttt{FeynCalc} using \texttt{FeynHelpers}
\cite{Shtabovenko_2017}.
 
The results for the considered one-loop contributions to the cross section as
well as for the \(2 \rightarrow 3\) cross sections have the general structure
\begin{equation} \label{X}
    \begin{aligned}
      \sigma_{k} = \sigma_{k}^{1/\epsilon_{\mathrm{IR}}^{2}} + \sigma_{k}^{1/\epsilon_{\mathrm{IR}}}
      + \Delta_{\mathrm{IR}} \, \sigma_{k}^{1/\epsilon_{\mathrm{IR}}} + \sigma_{k}^{\mathrm{fin}}
      + \Delta_{\mathrm{UV}} \, \sigma_{k}^{\mathrm{fin}} + \Delta_{\mathrm{IR}} \, \sigma_{k}^{\mathrm{fin}}
      + \pazocal{O} \left( \epsilon \right),
    \end{aligned}
\end{equation}
\noindent
where \(\Delta_{\mathrm{UV}}\) and \(\Delta_{\mathrm{IR}}\) denote new quantum
corrections arising from evanescent interactions cancelling UV- and
IR-divergences\footnote{The IR-divergences are  regularised dimensionally, not with a
small mass as IR regulator.}, respectively.
  
The tree-level cross section (in four dimensions) is
\begin{equation} \label{tree}
    \begin{aligned}
      \sigma_{\mathrm{tree}}\left(e e \rightarrow \mu \mu\right) = \frac{e^{4}}{12 \, \pi \, s}
      + \frac{y_{e}^{2} \, y_{\mu}^{2}}{16 \, \pi \, s}.
    \end{aligned}
\end{equation}
The one-loop contribution to the cross section, containing only muon vertex corrections,
is given in (\ref{Xloop1}) to (\ref{Xloop6})
in the \(\overline{\mathrm{MS}}\)-scheme and having used  decomposition
(\ref{X}).

The tree-level cross sections of muon production with real photon and real dilaton
emission, i.e. for \(e^{-} \, e^{+} \rightarrow \mu^{-} \, \mu^{+} \, \gamma\) and
\(e^{-} \, e^{+} \rightarrow \mu^{-} \, \mu^{+} \, \mathfrak{D}\), respectively,
arranged as in (\ref{X}), are provided in Appendix~\ref{D} as well.

The total cross section, considering virtual and real corrections only to the muon vertex,
has a form given in (\ref{XtotalD}).
The corresponding one-loop contribution to this total cross section, 
in the \(\overline{\mathrm{MS}}\)-scheme 
may be written as
\begin{equation} \label{s1}
    \begin{aligned}
      \sigma_{\mathrm{total, 1L} \mu}\left(e e \rightarrow \mu \mu\right) &=
      \sigma_{\mathrm{total, 1L} \mu}^{\mathrm{old}}\left(e e \rightarrow \mu \mu\right)
      + \Delta_{\mathrm{UV}} \, \sigma_{\mathrm{total, 1L} \mu}\left(e e \rightarrow \mu \mu\right)
    \end{aligned}
\end{equation}
\noindent
with the ``standard''\footnote{This is the usual result
found without imposing scale invariance at quantum level (QED with dilaton Yukawa couplings but employing DR instead of SIR).} one-loop contribution
\begin{equation} \label{s2}
    \begin{aligned}
      \sigma_{\mathrm{total, 1L} \mu}^{\mathrm{old}}\left(e e \rightarrow \mu \mu\right)
      &= \frac{1}{512 \, \pi^{3} \, s} \, \Big( 8 \, e^{6} - 4 \, e^{4} \, y_{\mu}^{2}
      + 34 \, e^{2} \, y_{e}^{2} \, y_{\mu}^{2} - 17 \, y_{e}^{2} \, y_{\mu}^{4} \Big)\\
      &- \frac{1}{256 \, \pi^{3} \, s} \, \Big[ 6 \, e^{2} \, y_{e}^{2} \, y_{\mu}^{2}
      - 3 \, y_{e}^{2} \, y_{\mu}^{4} \Big] \log\left(\frac{s}{\mu_{0}^{2}}\right)
    \end{aligned}
\end{equation}
\noindent
and the new, finite quantum correction that emerged from UV-divergences
\begin{equation} \label{s3}
    \begin{aligned}
      \Delta_{\mathrm{UV}} \, \sigma_{\mathrm{total, 1L} \mu}\left(e e \rightarrow \mu \mu\right)
      &= - \frac{1}{128 \, \pi^{3} \, s} \, y_{e}^{2} \, y_{\mu}^{2} \, \big( y_{e}^{2} + 4 \, y_{\mu}^{2} + y_{\tau}^{2} \big).
    \end{aligned}
\end{equation}
\noindent
with $s$ being the center-of-mass energy.
Thus, the considered one-loop contribution to cross section  (\ref{s1}),
that contains virtual and real corrections, is UV- \emph{and} IR-finite, as expected.
  Therefore, the IR poles do cancel in the SIR scheme. We can summarise the main results of this section
  as follows:
  
    \renewcommand{\labelenumi}{(\roman{enumi})}
    \begin{enumerate}
    \item The one-loop contribution (\ref{s1})
      contains the regular contribution (\ref{s2})
      that would  also be obtained in the usual massless QED with an additional
      scalar field coupling to the fermions,  and an additional contribution
      (\ref{s3})      which is a new quantum correction (from the UV sector)
      due to evanescent interactions. 

    \item The one-loop contribution (\ref{s1})
      to the cross section of \(e^{-} \, e^{+} \rightarrow \mu^{-} \, \mu^{+}\) scattering
      is IR-finite after summing over the contributing final state real emissions. Thus, all
      IR-divergences, even the new ones, cf. (\ref{Xloop3}),
      emerging as a result of evanescent interactions (due to quantum
      scale invariance), cancel after summing over the virtual and real corrections to the
      cross section. This is an important consistency check of theories with quantum scale invariance.
      
    \item All new, {\it finite} quantum corrections to the cross section, arising from evanescent
      interactions  ($\propto \epsilon^n$) that cancel IR poles, see (\ref{Xloop6}),
      cancel as well. Therefore,       quantum corrections due to the  UV poles given in eq.(\ref{Xloop5}),
      are the  only new finite contributions to the cross section that remain,
      as shown in (\ref{s3}). The essential reason for this
      cancellation is that the standard IR poles cancel between real
      and virtual diagrams of the same orders in all coupling
      constants. 
      The new IR quantum corrections then arise from the same
      evanescent interactions $\propto \epsilon^n$ in both real and virtual contributions
      hitting $1/\epsilon^m$ IR poles; hence they cancel
      simultaneously with the IR poles. 
      
      Results (ii) and (iii) are the main results of this section.

    \item The remaining new,  finite quantum correction to the cross section
      \(\Delta_{\mathrm{UV}} \, \sigma_{\mathrm{total, 1L} \mu}\), given in
      (\ref{s3}),      is an effect from the Yukawa sector. Hence, the gauge sector  does not give
      rise to new quantum corrections at one-loop, but these are expected
      at the two-loop level.

    \item Note that the new quantum correction (\ref{s3})
      is suppressed by Yukawa couplings to power \(6\), i.e. it
      is suppressed by $w^6$, where $w$ is the dilaton vev (using \(y_{f} = m_{f}/w\)).

    \item
      Since 
      \(\mu_{0} = \mu(\langle \sigma \rangle) \equiv \mu(w) = z \, w^{\frac{1}{1 - \epsilon}}\) and
      supposing that the vev of the Dilaton \(\langle \sigma \rangle \equiv w\) scales in the same
      way as  \(\sqrt{s}\), it can be seen that the tree-level cross section
      (\ref{tree}) and the 1-loop contribution
      (\ref{s1}) scale as \(\sim 1/s\), as expected on dimensional arguments for the cross section.
    \end{enumerate}


\section{Conclusions}

We studied scale invariance at the quantum level in (scale invariant) gauge theories that have
both UV and IR divergences. The interest in quantum (global) scale invariant
  theories is that they can naturally preserve an initial,  classical
  hierarchy of scales generated by fields vev's; this hierarchy could be
  arranged e.g. by one initial classical
  fine tuning of dimensionless coupling constants. More generally, quantum scale symmetry
  becomes necessary in gauge theories of scale invariance that include Einstein gravity.
Quantum scale invariance is achieved by implementing a scale invariant
regularisation and renormalisation,
in which the traditional subtraction scale is generated (dynamically)
by the vev of the dilaton, the Goldstone of the (global) scale symmetry.
  The vev of this field can be regarded, in a sense,  as the scale of ``new physics''.
In the limit of a large dilaton vev, when this field decouples, the quantum scale invariant
theory recovers the traditional quantum theory obtained without respecting scale invariance
at the quantum level (anomalous breaking). Above this scale, however, the two quantum theories can
differ in their UV completion, UV spectrum and quantum symmetry. 

We first checked the BRST invariance of quantum scale invariant theories. This can be done in two apparently
different formulations of scale invariant regularisations (SIR); we showed that the two approaches are equivalent and
that the BRST symmetry is maintained.

Further,  gauge theories have not only UV poles but also IR ones and one should check their fate
in quantum scale invariant theories using  the SIR scheme. We illustrated their analysis
in an Abelian gauge theory (quantum electrodynamics extended by a dilaton field).
While the UV poles do bring new quantum corrections (counterterms), finite or divergent,  beyond those
of the ``traditional'' approach, and which  we computed, it was shown that the IR poles do
not bring any new physics. The infrared  quantum corrections, {\it  both finite and divergent}, cancel out
similar to the way the IR poles themselves cancel in the traditional approach.
The cancellation is at the cross section level, after summing over the virtual and real
corrections. Therefore, the  ``evanescent'' interactions due to analytical continuation to $d$ dimensions
(SIR approach) do not affect the IR physics.
This was illustrated for the muon production cross section  \(e^{-} \, e^{+} \rightarrow \mu^{-} \, \mu^{+}\)
for the aforementioned Abelian theory. This result is a strong  consistency check of
quantum scale invariance in gauge theories.

\newpage
\section*{Appendix}

\def\theequation{A-\arabic{equation}}
\def\thesubsection{A}
\setcounter{equation}{0}
\def\thefigure{A-\arabic{figure}}
\def\thelabel{A}

\subsection{Feynman Rules} \label{Feynman-rules}

The Feynman rules for Lagrangian (\ref{QSIQED-taylor}) are shown below, for the
propagators:
\begin{figure}[!h]
\flushright{\includegraphics[scale=1]{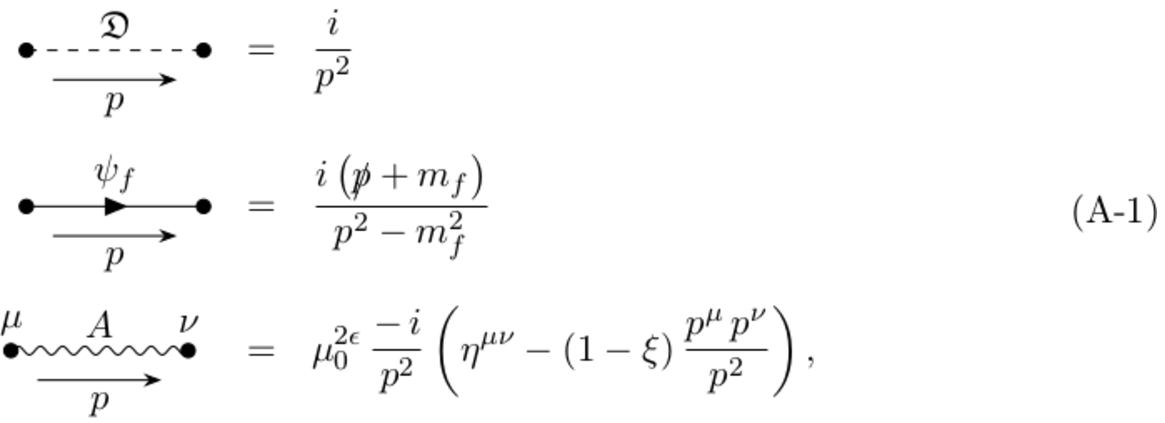}}
\end{figure}
\vspace{-0.6cm}

\noindent
vertices:

\begin{figure}[!h]
\flushright{\includegraphics[scale=1]{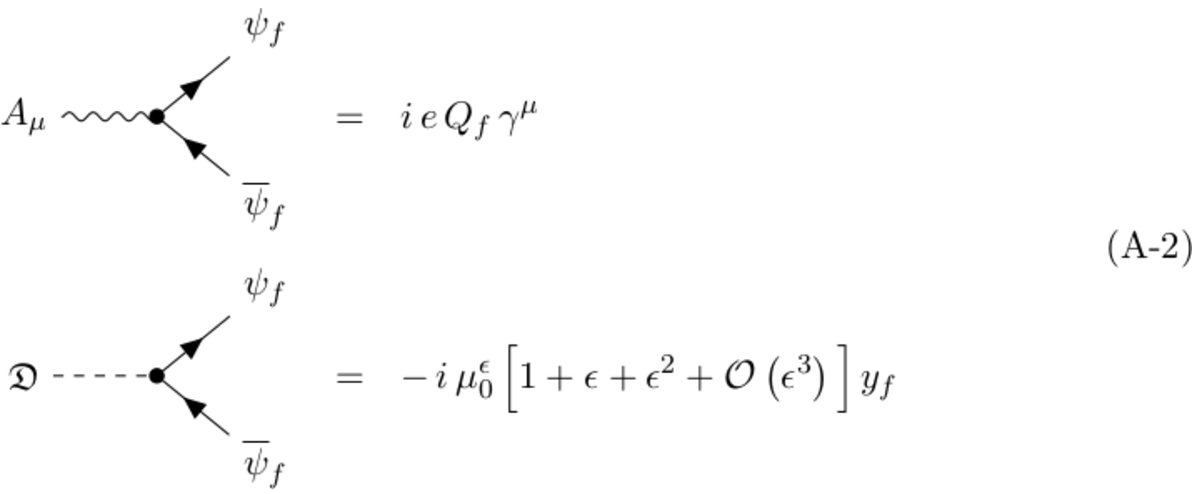}}
\end{figure}
\vspace{-0.5cm}

\noindent
and evanescent vertices:
\vspace{0.2cm}
\begin{figure}[!h]
\flushright{\includegraphics[scale=1]{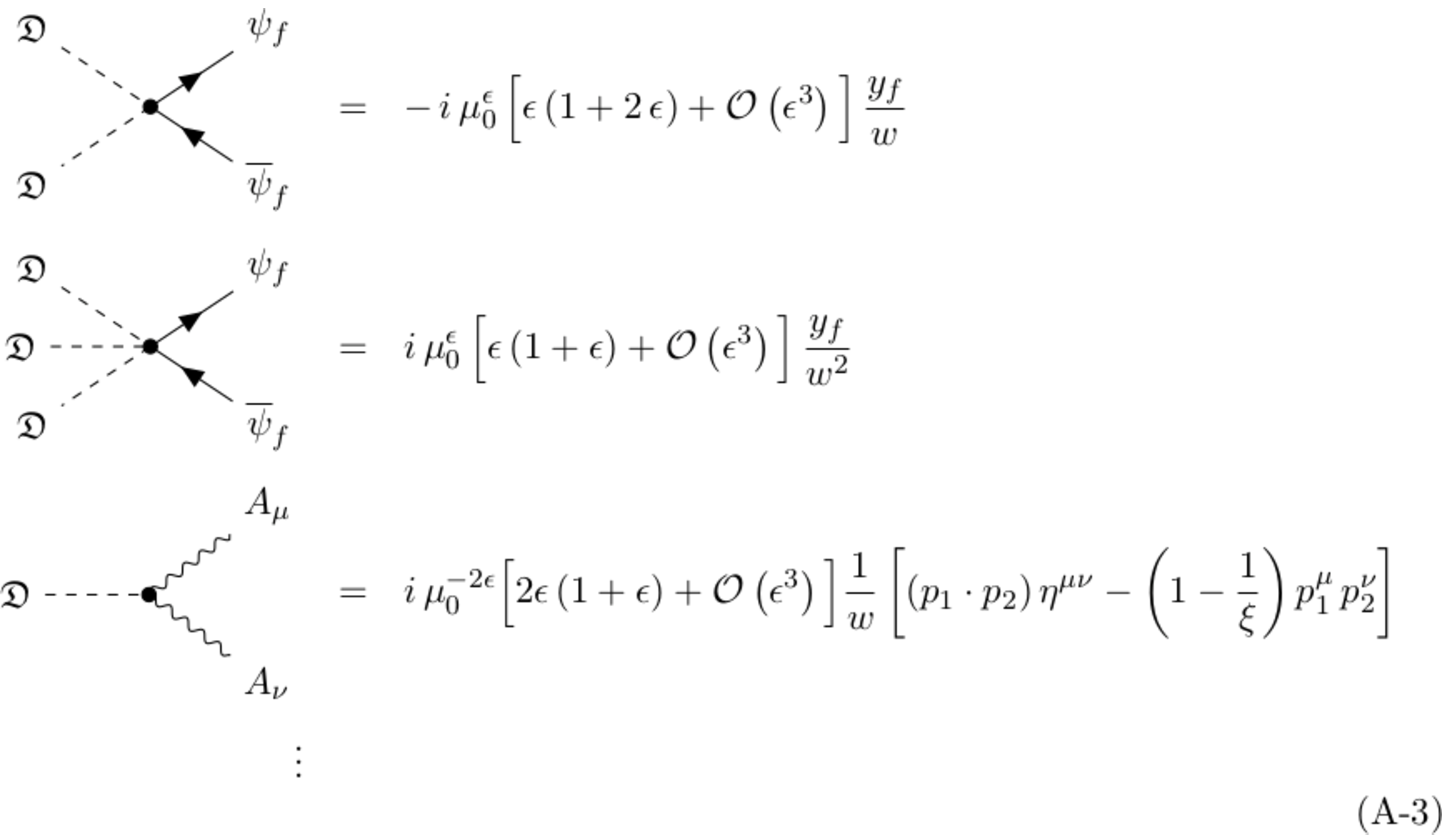}}
\end{figure}

\def\theequation{B-\arabic{equation}}
\def\thesubsection{B}
\setcounter{equation}{0}
\def\thefigure{B-\arabic{figure}}
\def\thelabel{B}

\subsection{Renormalisation of the Abelian case} \label{Renormalisation}

The quantum  scale invariant Abelian case  (QED + dilaton) of (\ref{QSIQED})
is renormalised as follows
\begin{equation} 
    \begin{aligned}
        \pazocal{L} &\longrightarrow \pazocal{L}^{(d)}_{0} = \pazocal{L}^{(d)}_{\mathrm{ren}} + \pazocal{L}^{(d)}_{\mathrm{ct}}\\
   		A &\longrightarrow A_{0} = \sqrt{Z_{A}} \, A\\
   		\psi_{f} &\longrightarrow \psi_{f, 0} = \sqrt{Z_{\psi_{f}}} \, \psi_{f}\\
        \sigma &\longrightarrow \sigma_{0} = \sqrt{Z_{\sigma}} \, \sigma\\
        e &\longrightarrow e_{0} = \mu^{\epsilon}\Big(\sqrt{Z_{\sigma}} \, \sigma\Big) \, Z_{e} \, e\\
        y_{f} &\longrightarrow y_{f, 0} = \mu^{\epsilon}\Big(\sqrt{Z_{\sigma}} \, \sigma\Big) \, Z_{y_{f}} \, y_{f}\\
        \lambda &\longrightarrow \lambda_{0} = \mu^{2 \epsilon}\Big(\sqrt{Z_{\sigma}} \, \sigma\Big) \, Z_{\lambda} \, \lambda\\
        \xi &\longrightarrow \xi_{0} = Z_{\xi} \, \xi
    \end{aligned}
\end{equation}
where $Z_{\xi} = Z_{A}$ due to the Ward identity. Despite setting
$\lambda \equiv 0$ at tree-level, as done in Eq.\ (\ref{QSIQED}),
the associated counterterm $\delta\lambda$ must be taken into account
for the complete one-loop renormalisation of the considered model.\footnote{
Note that $\delta\lambda$ does not contribute to the one-loop
muon production considered in Section \ref{3}.}

The one-loop counterterms\footnote{These counterterms are scale invariant.} of (\ref{QSIQED})
in the \(\overline{\mathrm{MS}}\)-scheme are generated by the  poles below
\begin{equation} 
    \begin{aligned}
        \delta Z_{\psi_{f}} &= - \frac{1}{16 \, \pi^{2}} \left[e^{2} + \frac{y_{f}}{2}\right] \frac{1}{\epsilon},
        \hspace{0.25cm} & \hspace{0.25cm}
        \delta Z_{y_{f}} &= \frac{1}{16 \, \pi^{2}} \left[ \frac{3}{2} \, y_{f}^{2} + \sum_{l} \, y_{l}^{2} - 3 \, e^{2} \right] \frac{1}{\epsilon},\\
        \delta Z_{A} &= - \frac{1}{16 \, \pi^{2}} \, \frac{4 \, N_{f} \, e^{2}}{3} \, \frac{1}{\epsilon},
        \hspace{0.25cm} & \hspace{0.25cm}
        \delta Z_{e} &= - \frac{1}{2} \, \delta Z_{A} = \frac{1}{16 \, \pi^{2}} \, \frac{2 \, N_{f} \, e^{2}}{3} \, \frac{1}{\epsilon},\\
        \delta Z_{\sigma} &= - \frac{1}{16 \, \pi^{2}} \, 2 \, \sum_{l} \, y_{l}^{2} \, \frac{1}{\epsilon},
        \hspace{0.25cm} & \hspace{0.25cm}
        \delta \lambda &= - \frac{1}{16 \, \pi^{2}} \, 24 \, \sum_{l} \, y_{l}^{4} \, \frac{1}{\epsilon},
    \end{aligned}
\end{equation}
where $f,l \in \{e^{-}, \mu^{-}, \tau^{-}\}$.

Note the non-vanishing wave function renormalisation $\delta Z_{\sigma}$. In contrast to the two-scalar model,
discussed in \cite{Ghilencea_2016, Ghilencea_2016_2Loop, kugo2020necessity, Lalak_2018, Olszewski_2017, Matthias_2021}, where the scalar field wave function renormalisation vanishes at the one-loop level, here the renormalisation of $\sigma$ in $\mu(\sigma)$ needs to be taken into account as it can give rise to new finite quantum corrections of the form $\epsilon \, \delta Z_{\sigma}$.\footnote{This is seen after expanding the Lagrangian w.r.t. $\epsilon$, $w$ and $\hbar$ for a given loop order.} Indeed, such a correction contributes to the new finite quantum correction of the one-loop cross section given in (\ref{s3}). In particular,
\begin{equation}
    \begin{aligned}
      \Delta_{\mathrm{UV}} \, \sigma_{\mathrm{total, 1L} \mu}\left(e e \rightarrow \mu \mu\right)
      &= - \frac{3}{128 \, \pi^{3} \, s} \, y_{e}^{2} \, y_{\mu}^{4} + \sigma^{\mathrm{Yuk}}_{\mathrm{tree}}\left(e e \rightarrow \mu \mu\right) \epsilon \, \delta Z_{\sigma}\\
      &= - \frac{3}{128 \, \pi^{3} \, s} \, y_{e}^{2} \, y_{\mu}^{4} + \frac{y_{e}^{2} \, y_{\mu}^{2}}{16 \pi s} \, \epsilon \, \delta Z_{\sigma}\\
      &= - \frac{1}{128 \, \pi^{3} \, s} \, y_{e}^{2} \, y_{\mu}^{2} \, \big( y_{e}^{2} + 4 \, y_{\mu}^{2} + y_{\tau}^{2} \big),
    \end{aligned}
\end{equation}
with $\sigma^{\mathrm{Yuk}}_{\mathrm{tree}}(e e \rightarrow \mu \mu) = y_{e}^{2} \, y_{\mu}^{2}/(16 \pi s)$ being the tree-level contribution of the Yukawa sector.

\def\theequation{C-\arabic{equation}}
\def\thesubsection{C}
\setcounter{equation}{0}
\def\thefigure{C-\arabic{figure}}
\def\thelabel{C}

\subsection{Feynman diagrams for muon production}\label{Feynman}

In the theory considered in
(\ref{QSIQED-taylor}), there are two Feynman diagrams
contributing to the scattering process \(e^{-} \, e^{+} \longrightarrow \mu^{-} \, \mu^{+}\) at tree-level:
\medskip
\begin{figure}[!h]
\flushright{\includegraphics[scale=1]{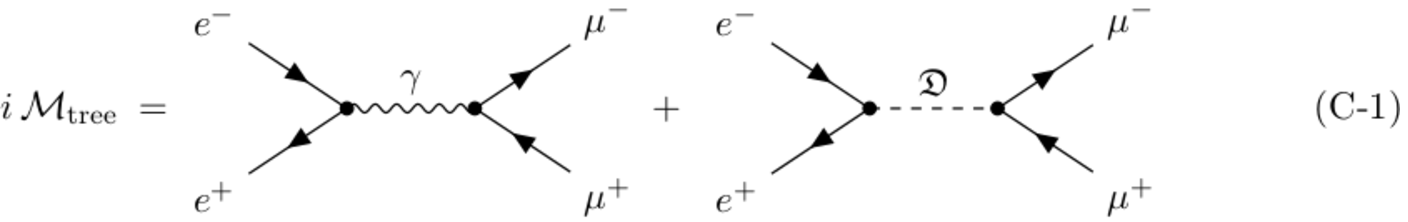}}
\end{figure}

\smallskip
\noindent
At the one-loop level, there are ten Feynman diagrams containing a one-loop muon vertex correction, four of them with a photon mediator as illustrated in (C-2)
  \medskip
  \begin{figure}[!h]
\flushright{\includegraphics[scale=1]{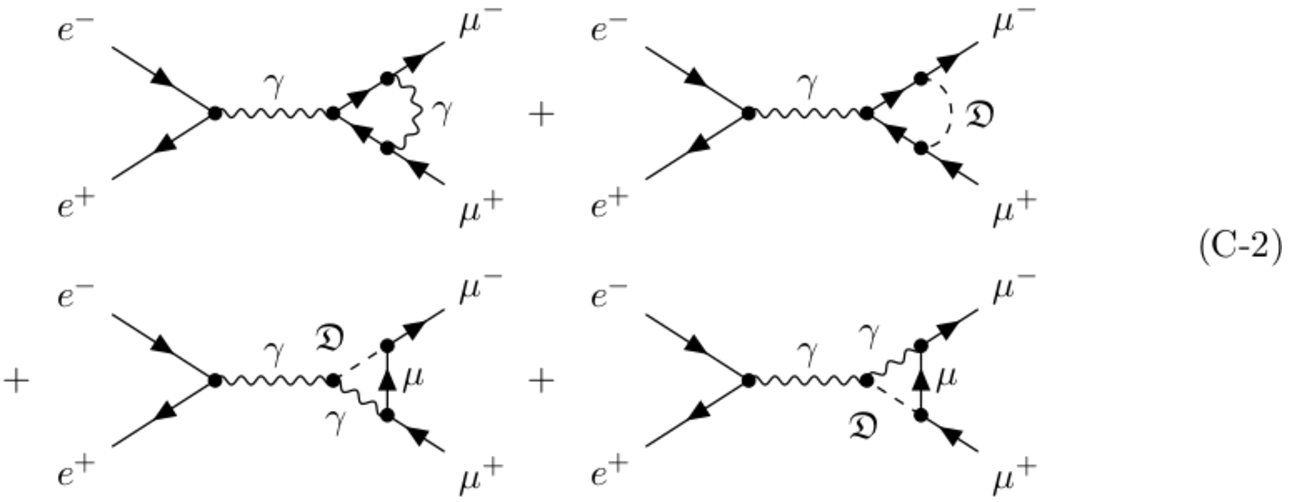}}
\end{figure}
  
\medskip
\noindent
and the other six with a dilaton mediator as shown below in (C-3)
  \bigskip
\begin{figure}[!h] %
  \flushright{\includegraphics[scale=1]{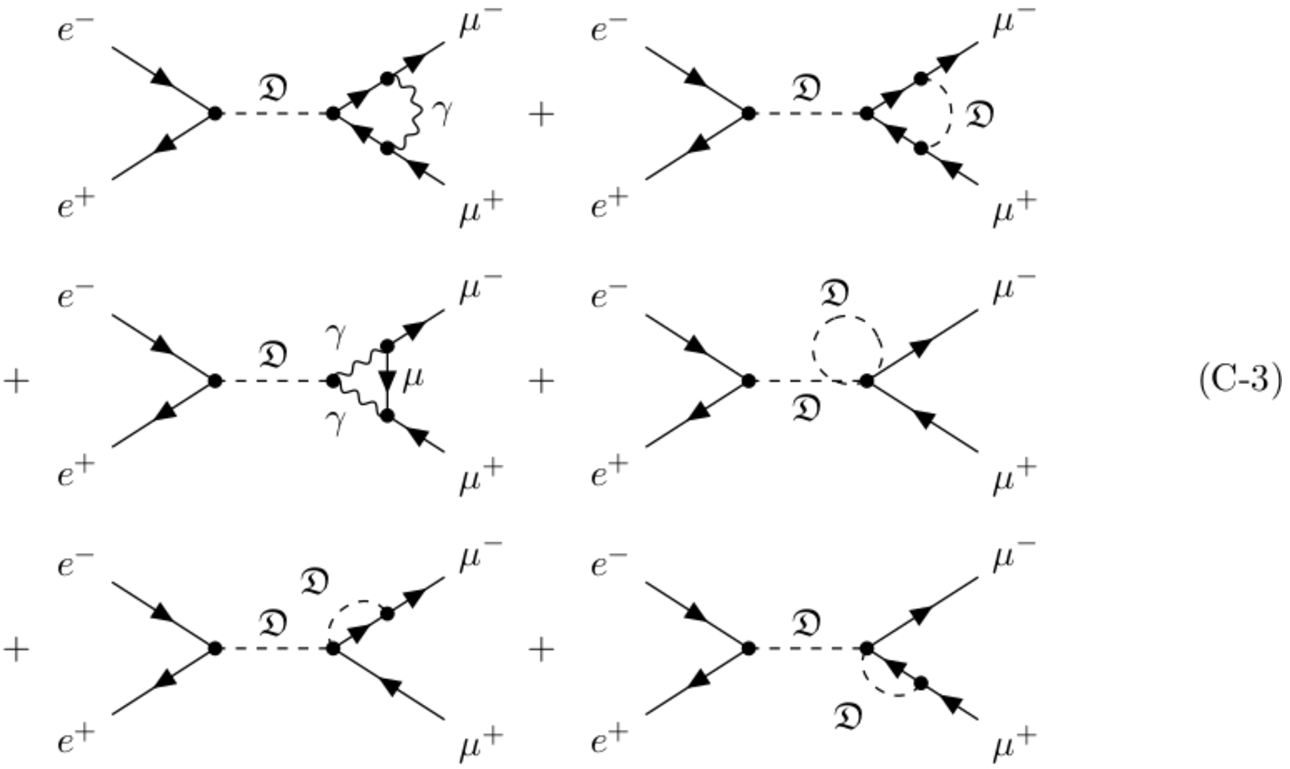}}
\end{figure}

\vspace{2cm}
\noindent
For the scattering process \(e^{-} \, e^{+} \longrightarrow \mu^{-} \, \mu^{+} \, \gamma\)
there are six tree-level Feynman diagrams of which three are photon mediated as in (C-4)
\bigskip
\begin{figure}[!h] %
  \flushright{ \includegraphics[scale=1]{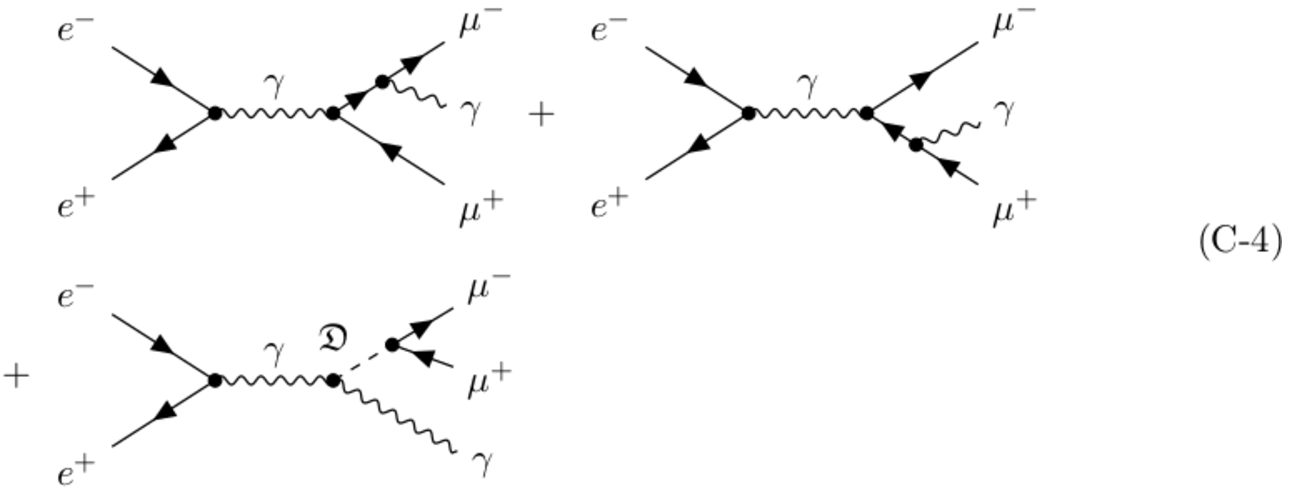}}
\end{figure}

\medskip
 \noindent
 and the remaining three are dilaton mediated, as illustrated in (C-5)
 \bigskip
 \begin{figure}[!h] %
  \flushright{ \includegraphics[scale=1]{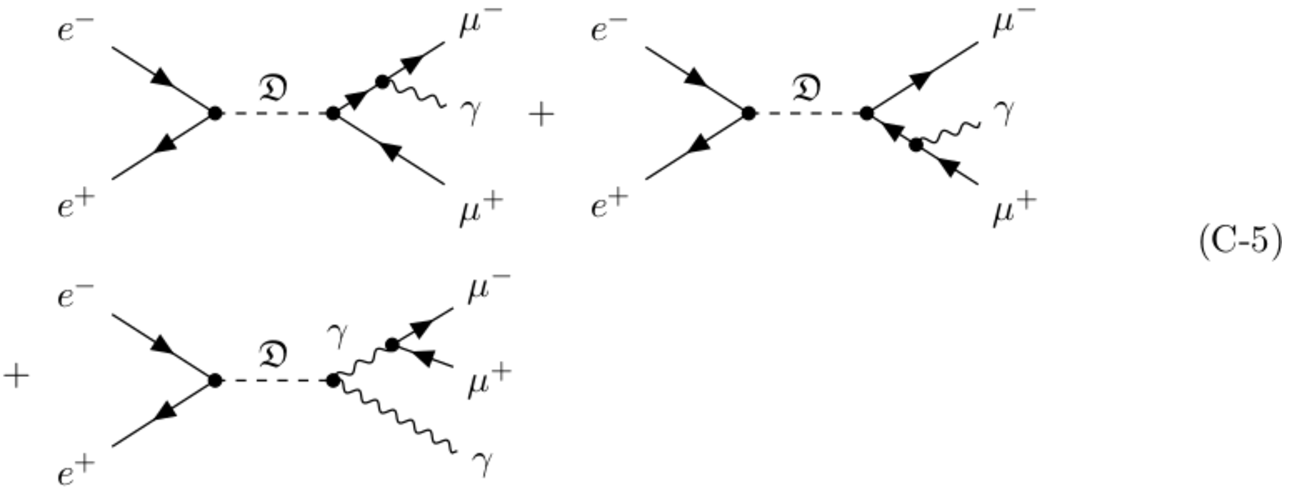}}
 \end{figure}

 \medskip
 \noindent{Additionally, for the scattering process \(e^{-} \, e^{+} \longrightarrow \mu^{-} \, \mu^{+} \, \mathfrak{D}\) there are six tree-level Feynman diagrams.}
 Three of them are photon mediated as shown in (C-6)
 \bigskip \bigskip
\begin{figure}[!h] %
 \flushright{  \includegraphics[scale=1]{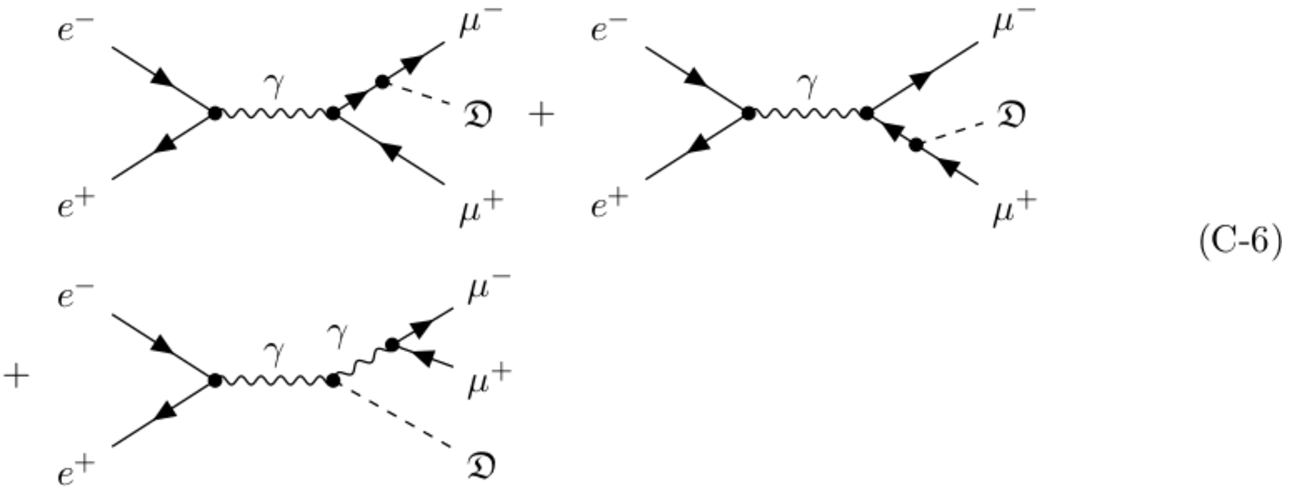}}
 \end{figure}

\vspace{0.5cm}
\noindent
whereas the remaining three of them are dilaton mediated, shown below in (C-7)
  \bigskip

\begin{figure}[!h] %
\flushright{\includegraphics[scale=1]{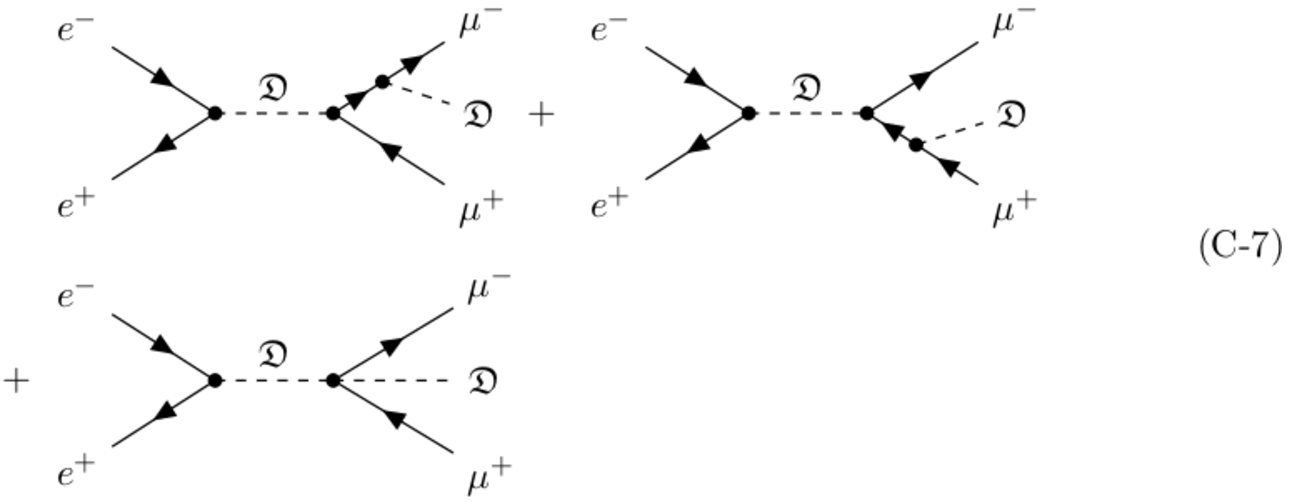}}
\vspace{1.5cm}
 \end{figure}

\def\theequation{D-\arabic{equation}}
\def\thesubsection{D}
\setcounter{equation}{0}
\def\thefigure{D-\arabic{figure}}
\def\thelabel{D}

\subsection{Muon production: cross sections} \label{D}

In the following, explicit results for the cross section of muon production are provided at
the one-loop level including virtual and real corrections to the muon vertex only.\\

\noindent{ \textbf{ { (i) Tree-level: } } }\\

\noindent{The tree-level cross section, in \(4\) dimensions, is provided by}
\begin{equation} \label{Xtree}
    \begin{aligned}
        \sigma_{\mathrm{tree}}\left(e e \rightarrow \mu \mu\right) = \frac{e^{4}}{12 \, \pi \, s} + \frac{y_{e}^{2} \, y_{\mu}^{2}}{16 \, \pi \, s}
    \end{aligned}
\end{equation}\\

\noindent{ \textbf{ { (ii) 1-loop: } } }\\

\noindent{Using (\ref{X}), the 1-loop cross section for \(e^{-} \, e^{+} \longrightarrow \mu^{-} \, \mu^{+}\), containing only 1-loop muon vertex corrections, in the \(\overline{\mathrm{MS}}\)-scheme, is
the sum of the following contributions}
\begin{equation} \label{Xloop1}
    \begin{aligned}
        \sigma_{\mathrm{1L}, \mu}^{1/\epsilon_{\mathrm{IR}}^{2}}\left(e e \rightarrow \mu \mu\right) = - \frac{\mu_{0}^{2 \epsilon}}{192 \, \pi^{3} \, s} \Big( 4 \, e^{6} + 3 \, e^{2} \, y_{e}^{2} \, y_{\mu}^{2} \Big) \frac{1}{\epsilon_{\mathrm{IR}}^{2}}
    \end{aligned}
\end{equation}
\begin{equation} \label{Xloop2}
    \begin{aligned}
        \sigma_{\mathrm{1L}, \mu}^{1/\epsilon_{\mathrm{IR}}}\left(e e \rightarrow \mu \mu\right) = &- \frac{\mu_{0}^{2 \epsilon}}{2304 \, \pi^{3} \, s} \bigg[ 104 \, e^{6} - 12 \, e^{4} \, y_{\mu}^{2} + 126 \, e^{2} \, y_{e}^{2} \, y_{\mu}^{2} - 9 \, y_{e}^{2} \, y_{\mu}^{4}\\
        &\hspace{2.2cm} - 24 \, \Big( 4 \, e^{6} + 3 \, e^{2} \, y_{e}^{2} \, y_{\mu}^{2} \Big) \log\left(\frac{s}{\mu_{0}^{2}}\right) \bigg] \frac{1}{\epsilon_{\mathrm{IR}}}
    \end{aligned}
\end{equation}
\begin{equation} \label{Xloop3}
    \begin{aligned}
        \Delta_{\mathrm{IR}} \, \sigma_{\mathrm{1L}, \mu}^{1/\epsilon_{\mathrm{IR}}}\left(e e \rightarrow \mu \mu\right) = &- \frac{\mu_{0}^{2 \epsilon}}{16 \, \pi^{3} \, s} \, e^{2} \, y_{e}^{2} \, y_{\mu}^{2} \, \frac{1}{\epsilon_{\mathrm{IR}}}
    \end{aligned}
\end{equation}
\begin{equation} \label{Xloop4}
    \begin{aligned}
        \sigma_{\mathrm{1L}, \mu}^{\mathrm{fin}}\left(e e \rightarrow \mu \mu\right) &= \frac{\mu_{0}^{2 \epsilon}}{3456 \, \pi^{3} \, s} \bigg[ \Big( 60 \, \pi^{2} - 464 \Big) \, e^{6} + 30 \, e^{4} \, y_{\mu}^{2}\\
        &\hspace{2.1cm} + 9 \, \Big( 5 \, \pi^{2} - 48 \Big) \, e^{2} \, y_{e}^{2} \, y_{\mu}^{2} - 27 \, y_{e}^{2} \, y_{\mu}^{4} \bigg]\\
        &+ \frac{\mu_{0}^{2 \epsilon}}{2304 \, \pi^{3} \, s} \bigg[ 208 \, e^{6} - 24 \, e^{4} \, y_{\mu}^{2} + 198 \, e^{2} \, y_{e}^{2} \, y_{\mu}^{2} + 9 \, y_{e}^{2} \, y_{\mu}^{4} \bigg] \log\left(\frac{s}{\mu_{0}^{2}}\right)\\
        &- \frac{\mu_{0}^{2 \epsilon}}{96 \, \pi^{3} \, s} \bigg[ 4 \, e^{6} + 3 \, e^{2} \, y_{e}^{2} \, y_{\mu}^{2} \bigg] \log^{2}\left(\frac{s}{\mu_{0}^{2}}\right)
    \end{aligned}
\end{equation}
\begin{equation} \label{Xloop5}
    \begin{aligned}
        \Delta_{\mathrm{UV}} \, \sigma_{\mathrm{1L}, \mu}^{\mathrm{fin}}\left(e e \rightarrow \mu \mu\right) = &- \frac{\mu_{0}^{2 \epsilon}}{128 \, \pi^{3} \, s} \, y_{e}^{2} \, y_{\mu}^{2} \, \big( y_{e}^{2} + 4 \, y_{\mu}^{2} + y_{\tau}^{2} \big)
    \end{aligned}
\end{equation}
\begin{equation} \label{Xloop6}
    \begin{aligned}
        \Delta_{\mathrm{IR}} \, \sigma_{\mathrm{1L}, \mu}^{\mathrm{fin}}\left(e e \rightarrow \mu \mu\right) &= \frac{\mu_{0}^{2 \epsilon}}{384 \, \pi^{3} \, s} \bigg[ 4 \, e^{4} \, y_{\mu}^{2} - 84 \, e^{2} \, y_{e}^{2} \, y_{\mu}^{2} + 9 \, y_{e}^{2} \, y_{\mu}^{4} \bigg]\\
        &+ \frac{\mu_{0}^{2 \epsilon}}{8 \, \pi^{3} \, s} \, e^{2} \, y_{e}^{2} \, y_{\mu}^{2} \, \log\left(\frac{s}{\mu_{0}^{2}}\right) - \frac{5 \, \mu_{0}^{2 \epsilon}}{32 \, \pi^{3} \, s} \, e^{2} \, y_{e}^{2} \, y_{\mu}^{2}
    \end{aligned}
\end{equation}\\

\bigskip
\noindent{ \textbf{ { (iii) Real photon emission: } } }\\

\noindent
The contributions to the
tree-level cross section for \(e^{-} \, e^{+} \longrightarrow \mu^{-} \, \mu^{+} \, \gamma\),
arranged as in (\ref{X}), are
\begin{equation} \label{RE}
    \begin{aligned}
        \sigma_{\mathrm{tree}}^{1/\epsilon_{\mathrm{IR}}^{2}}\left(e e \rightarrow \mu \mu \gamma\right) = \frac{\mu_{0}^{2 \epsilon}}{192 \, \pi^{3} \, s} \Big( 4 \, e^{6} + 3 \, e^{2} \, y_{e}^{2} \, y_{\mu}^{2} \Big) \frac{1}{\epsilon_{\mathrm{IR}}^{2}}
    \end{aligned}
\end{equation}
\begin{equation} 
    \begin{aligned}
        \sigma_{\mathrm{tree}}^{1/\epsilon_{\mathrm{IR}}}\left(e e \rightarrow \mu \mu \gamma\right) &= \frac{\mu_{0}^{2 \epsilon}}{1152 \, \pi^{3} \, s} \bigg[ 52 \, e^{6} + 63 \, e^{2} \, y_{e}^{2} \, y_{\mu}^{2}\\
        &\hspace{2.2cm} - 12 \, \Big( 4 \, e^{6} + 3 \, e^{2} \, y_{e}^{2} \, y_{\mu}^{2} \Big) \log\left(\frac{s}{\mu_{0}^{2}}\right) \bigg] \frac{1}{\epsilon_{\mathrm{IR}}}
    \end{aligned}
\end{equation}
\begin{equation} 
    \begin{aligned}
        \Delta_{\mathrm{IR}} \, \sigma_{\mathrm{tree}}^{1/\epsilon_{\mathrm{IR}}}\left(e e \rightarrow \mu \mu \gamma\right) = \frac{\mu_{0}^{2 \epsilon}}{16 \, \pi^{3} \, s} \, e^{2} \, y_{e}^{2} \, y_{\mu}^{2} \, \frac{1}{\epsilon_{\mathrm{IR}}}
    \end{aligned}
\end{equation}
\begin{equation} 
    \begin{aligned}
        \sigma_{\mathrm{tree}}^{\mathrm{fin}}\left(e e \rightarrow \mu \mu \gamma\right) = &- \frac{\mu_{0}^{2 \epsilon}}{6912 \, \pi^{3} \, s} \bigg[ 4 \, \Big( 30 \, \pi^{2} - 259 \Big) \, e^{6} + 9 \, \Big( 10 \, \pi^{2} - 147 \Big) \, e^{2} \, y_{e}^{2} \, y_{\mu}^{2} \bigg]\\
        &- \frac{\mu_{0}^{2 \epsilon}}{576 \, \pi^{3} \, s} \bigg[ 52 \, e^{6} + 63 \, e^{2} \, y_{e}^{2} \, y_{\mu}^{2} \bigg] \log\left(\frac{s}{\mu_{0}^{2}}\right)\\
        &+ \frac{\mu_{0}^{2 \epsilon}}{96 \, \pi^{3} \, s} \bigg[ 4 \, e^{6} + 3 \, e^{2} \, y_{e}^{2} \, y_{\mu}^{2} \bigg] \log^{2}\left(\frac{s}{\mu_{0}^{2}}\right)
    \end{aligned}
\end{equation}
\begin{equation} 
    \begin{aligned}
        \Delta_{\mathrm{UV}} \, \sigma_{\mathrm{tree}}^{\mathrm{fin}}\left(e e \rightarrow \mu \mu \gamma\right) = 0
    \end{aligned}
\end{equation}
\begin{equation} 
    \begin{aligned}
        \Delta_{\mathrm{IR}} \, \sigma_{\mathrm{tree}}^{\mathrm{fin}}\left(e e \rightarrow \mu \mu \gamma\right) &= \frac{7 \, \mu_{0}^{2 \epsilon}}{32 \, \pi^{3} \, s} \, e^{2} \, y_{e}^{2} \, y_{\mu}^{2} - \frac{\mu_{0}^{2 \epsilon}}{8 \, \pi^{3} \, s} \, e^{2} \, y_{e}^{2} \, y_{\mu}^{2} \, \log\left(\frac{s}{\mu_{0}^{2}}\right)\\
        &+ \frac{5 \, \mu_{0}^{2 \epsilon}}{32 \, \pi^{3} \, s} \, e^{2} \, y_{e}^{2} \, y_{\mu}^{2}
    \end{aligned}
  \end{equation}\\

\noindent{ \textbf{ { (iv) Real dilaton emission: } } }

\bigskip
\noindent{Similarly, the   tree-level cross section for \(e^{-} \, e^{+} \longrightarrow \mu^{-} \,
  \mu^{+} \, \mathfrak{D}\) is the sum of the following  terms}
\begin{equation} 
    \begin{aligned}
        \sigma_{\mathrm{tree}}^{1/\epsilon_{\mathrm{IR}}^{2}}\left(e e \rightarrow \mu \mu \mathfrak{D}\right) = 0
    \end{aligned}
\end{equation}
\begin{equation} 
    \begin{aligned}
        \sigma_{\mathrm{tree}}^{1/\epsilon_{\mathrm{IR}}}\left(e e \rightarrow \mu \mu \mathfrak{D}\right) &= - \frac{\mu_{0}^{2 \epsilon}}{768 \, \pi^{3} \, s} \Big( 4 \, e^{4} \, y_{\mu}^{2} + 3 \, y_{e}^{2} \, y_{\mu}^{4} \Big) \frac{1}{\epsilon_{\mathrm{IR}}}
    \end{aligned}
\end{equation}
\begin{equation} 
    \begin{aligned}
        \Delta_{\mathrm{IR}} \, \sigma_{\mathrm{tree}}^{1/\epsilon_{\mathrm{IR}}}\left(e e \rightarrow \mu \mu \mathfrak{D}\right) = 0
    \end{aligned}
\end{equation}
\begin{equation} 
    \begin{aligned}
        \sigma_{\mathrm{tree}}^{\mathrm{fin}}\left(e e \rightarrow \mu \mu \mathfrak{D}\right) = &- \frac{\mu_{0}^{2 \epsilon}}{4608 \, \pi^{3} \, s} \Big( 76 \, e^{4} \, y_{\mu}^{2} + 117 \, y_{e}^{2} \, y_{\mu}^{4} \Big)\\
        &+ \frac{\mu_{0}^{2 \epsilon}}{384 \, \pi^{3} \, s} \bigg[ 4 \, e^{4} \, y_{\mu}^{2} + 3 \, y_{e}^{2} \, y_{\mu}^{4} \bigg] \log\left(\frac{s}{\mu_{0}^{2}}\right)\\
    \end{aligned}
\end{equation}
\begin{equation} 
    \begin{aligned}
        \Delta_{\mathrm{UV}} \, \sigma_{\mathrm{tree}}^{\mathrm{fin}}\left(e e \rightarrow \mu \mu \mathfrak{D}\right) = 0
    \end{aligned}
\end{equation}
\begin{equation} 
    \begin{aligned}
        \Delta_{\mathrm{IR}} \, \sigma_{\mathrm{tree}}^{\mathrm{fin}}\left(e e \rightarrow \mu \mu \mathfrak{D}\right) = - \frac{\mu_{0}^{2 \epsilon}}{384 \, \pi^{3} \, s} \Big( 4 \, e^{4} \, y_{\mu}^{2} + 9 \, y_{e}^{2} \, y_{\mu}^{4} \Big)
    \end{aligned}
\end{equation}

\bigskip
\noindent{ \textbf{ { (v) Total cross section: } } }

\bigskip
\noindent{Finally, the total cross section, considering only virtual and real corrections to the muon vertex, is given by}
\begin{equation} \label{XtotalD}
    \begin{aligned}
        \sigma_{\mathrm{total}, \mu}\left(e e \rightarrow \mu \mu\right) &= \sigma_{\mathrm{tree}}\left(e e \rightarrow \mu \mu\right) + \sigma_{\mathrm{1L}, \mu}\left(e e \rightarrow \mu \mu\right)\\
        &\hspace{0.42cm} + \sigma_{\mathrm{tree}}\left(e e \rightarrow \mu \mu \gamma\right) + \sigma_{\mathrm{tree}}\left(e e \rightarrow \mu \mu \mathfrak{D}\right) + \pazocal{O}\left(\alpha_{i}^{4}\right)\\
        &= \sigma_{\mathrm{tree}}\left(e e \rightarrow \mu \mu\right) + \sigma_{\mathrm{total, 1L} \mu}\left(e e \rightarrow \mu \mu\right) + \pazocal{O}\left(\alpha_{i}^{4}\right)
    \end{aligned}
\end{equation}
\noindent{where \(\alpha_{i}\) are the fine structure constants for \(e\) and \(y_{i}\).}
The full tree-level cross section is to be found in (\ref{Xtree}), whereas the considered one-loop contribution in the \(\overline{\mathrm{MS}}\)-scheme and in \(4\) dimensions is given by
\begin{equation} 
    \begin{aligned}
        \sigma_{\mathrm{total, 1L} \mu}\left(e e \rightarrow \mu \mu\right) &= \sigma_{\mathrm{total, 1L} \mu}^{\mathrm{old}}\left(e e \rightarrow \mu \mu\right) + \Delta_{\mathrm{UV}} \, \sigma_{\mathrm{total, 1L} \mu}\left(e e \rightarrow \mu \mu\right)
    \end{aligned}
\end{equation}
\noindent{with the ``standard'' one-loop contribution}
\begin{equation} 
    \begin{aligned}
        \sigma_{\mathrm{total, 1L} \mu}^{\mathrm{old}}\left(e e \rightarrow \mu \mu\right) &= \frac{1}{512 \, \pi^{3} \, s} \, \Big( 8 \, e^{6} - 4 \, e^{4} \, y_{\mu}^{2} + 34 \, e^{2} \, y_{e}^{2} \, y_{\mu}^{2} - 17 \, y_{e}^{2} \, y_{\mu}^{4} \Big)\\
        &- \frac{1}{256 \, \pi^{3} \, s} \, \Big[ 6 \, e^{2} \, y_{e}^{2} \, y_{\mu}^{2} - 3 \, y_{e}^{2} \, y_{\mu}^{4} \Big] \log\left(\frac{s}{\mu_{0}^{2}}\right)
    \end{aligned}
\end{equation}
\noindent
and the new finite quantum correction that emerged from UV-divergences,
  quoted in eq.(\ref{s3})
\begin{equation} 
    \begin{aligned}
        \Delta_{\mathrm{UV}} \, \sigma_{\mathrm{total, 1L} \mu}\left(e e \rightarrow \mu \mu\right) &= - \frac{1}{128 \, \pi^{3} \, s} \, y_{e}^{2} \, y_{\mu}^{2} \, \big( y_{e}^{2} + 4 \, y_{\mu}^{2} + y_{\tau}^{2} \big).
    \end{aligned}
  \end{equation}

\bigskip\noindent
{\bf Acknowledgement:} The work of D.Ghilencea  was supported by\,a\,grant of\,the Romanian Ministry
of\,Education and Research CNCS-UEFISCDI project\,PN-III-P4-ID-PCE-2020-2255.


\end{document}